\documentclass[prb,twocolumn,showpacs,eqsecnum,unsortedaddress]{revtex4}
\usepackage{graphicx,epsfig}
\usepackage{amsmath,amssymb,mathrsfs}
\let\eps=\epsilon
\let\veps=\varepsilon
\let\up=\uparrow
\let\down=\downarrow 

\newcommand\avg[1]{\left\langle{\textstyle #1}\right\rangle}
\newcommand\ket[1]{\left|{\textstyle #1}\right\rangle}
\newcommand\bra[1]{\left\langle{\textstyle #1}\right|}
\newcommand\varN{\mathscr{N}}
\newcommand\varH{\mathscr{H}}
\newcommand\varS{\mathscr{S}}
\newcommand\half{\frac{1}{2}}

\newcommand\Kondo{\mathrm{Kondo}}

\newcommand\vtilH{\widetilde\varH}
\newcommand\tilV{\widetilde{V}}
\newcommand\bfS{\mathbf{S}}
\newcommand\bfT{\overline{\mathbf{T}}}
\newcommand\bfsigma{{\boldsymbol{\sigma}}}
\newcommand\bftau{{\boldsymbol{\tau}}}
\newcommand\tr{\mathrm{Tr}}
\newcommand\im{\mathrm{Im}}

\begin{document}

\title{Probing spin and orbital Kondo effects with a mesoscopic interferometer}

\author{Rosa L\'opez}
\author{David S\'anchez}
\affiliation{D\'epartement de Physique Th\'eorique,
  Universit\'e de Gen\`eve, CH-1211 Gen\`eve 4, Switzerland}

\author{Minchul Lee}
\author{Mahn-Soo Choi}
\affiliation{Department of Physics, Korea University, Seoul 136-701,
  Korea}
\author{Pascal Simon}
\affiliation{Laboratoire de Physique et Mod\'elisation des Milieux
  Condens\'es, CNRS et UJF, 38042 Grenoble, France}

\author{Karyn Le Hur}
\affiliation{D\'epartement de Physique and RQMP, Universit\'e de
  Sherbrooke, Sherbrooke, Qu\'ebec, Canada, J1K, 2R1}

\begin{abstract}
We investigate theoretically the transport properties
of a closed Aharonov-Bohm interferometer containing two quantum
dots in the strong coupling regime. We find two distinct physical 
scenarios depending on the strength of the interdot Coulomb interaction. 
When the interdot Coulomb interaction is negligible only spin 
fluctuations are important and each dot develops a Kondo resonance 
at the Fermi level independently of the applied
magnetic flux. The transport is characterized by the interference 
of these two independent Kondo resonances. On the contrary, for large 
interdot interaction, only one electron can be accommodated 
onto the double dot system. In this situation, not only the spin 
can fluctuate but also the orbital  degree of freedom (the pseudo-spin). 
As a result, we find different ground states depending on the value of the
applied flux. When $\phi=\pi$ (mod $2\pi$) ($\phi=2\pi\Phi/\Phi_0$, 
where $\Phi$ is applied flux, and $\Phi_0=h/e$ the flux quantum) 
the electronic transport can take place via simultaneous 
correlations in the spin and pseudo-spin sectors, leading to the 
highly symmetric SU(4) Kondo state. 
Nevertheless, we find situations with $\phi>0$ (mod $2\pi$) 
where the pseudo-spin quantum number is not conserved during 
tunneling events, giving rise to the common 
SU(2) Kondo state with an enhanced Kondo temperature. 
We investigate the crossover between both ground states 
and discuss possible experimental signatures of this physics 
as a function of the applied magnetic flux. 
\end{abstract}

\pacs{72.15.Qm, 73.63.Kv, 73.23.-b}
\maketitle

\section{Introduction}
Progressive advance in nanofabrication technology has
achieved the realization of tiny droplets of electrons termed
quantum dots (QD's) with a high-precision tunability of the 
transport parameters.~\cite{dot} One of the most exciting 
features of a QD is its ability to behave as a quantum 
impurity with spin 1/2.~\cite{gla88,ng88} At temperatures 
lower than the Kondo temperature ($T_K$), the localized 
spin becomes strongly correlated with the conduction
electrons and consequently is screened.~\cite{gla88,ng88,hew93} 
As a result of the increasing rate of scattering 
there arises a resonance at the Fermi energy ($E_F$) in the
density of states (DOS) of the QD.~\cite{hew93} The transmission 
through the quantum dot is then almost perfect. This is the so-called 
\emph{unitary limit} where conductance reaches $2e^2/h$.
~\cite{gol98,cro98,shm98} Among many of the advantages offered 
by QD-based devices, we highlight the possibility of studying the Kondo
effect out of equilibrium by applying a dc bias~\cite{noneq} or
a time-dependent potential.~\cite{acfield,time} 

A natural step forward is the understanding of the magnetic interactions
of two artificial Kondo impurities.~\cite{izu,geo99,ram00,eto01,
ros02,ros04,chan01,chan03,crai04}. The investigation 
of double QD's is mainly motivated by the possibility  of
their application as solid-state quantum bits, by using either spin or charge
degrees of freedom.~\cite{loss,chan03,crai04,dqdreview} When the two QD's are
interacting, the orbital degrees of freedom come into play as a pseudo-spin, as 
shown experimentally in Ref.~\onlinecite{Weiss,hol03}, which may give rise to exotic
physical scenarios.~\cite{Schon,Zar03} Thus, in a double QD it is possible to
tune appropriately the gate voltages in order to find two charge 
states almost degenerate. If the {\em inter}dot Coulomb interaction 
is large enough these two states are $\{n_{1}=1$, $n_{2}=0\}$ 
and $\{n_{1}=0$, $n_{2}=1\}$) where $n_{1(2)}=\langle 
\hat{n}_{1(2)}\rangle$  is the charge state in the dot
``$1$'' (``$2$''). This is one of the basic ingredients to observe Kondo
physics: the existence of degeneracy between two quantum states. 
The Kondo effect is then developed to its fullest extent in the
 pseudo-spin (orbital) sector. We define the pseudo-spin 
$\hat{T}$ as follows: it points along $+$($-$)$z$ when 
the electron is at the ``$1$'' (``$2$'') dot: $\hat{T}^{z}=(1/2)
(\hat{n}_{1}-\hat{n}_{2})$.  The pseudo-spin of the double
QD system can be $1/2$ or $-1/2$, which is quenched (screened) via higher
order tunneling processes producing the so-called \emph{orbital 
Kondo effect}.~\cite{Schon,Weiss,hol03} Other realizations of such exotic 
``orbital Kondo effects'' have been recently proposed in different 
QD-related structures as well.~\cite{Zar03,Karyn03a,Karyn03b}  
When the {\em intra}dot Coulomb energy for each dot is large, 
then each QD also behaves as an magnetic impurity and the 
conventional Kondo effect is also observed in the spin sector
($S^{z}=\pm 1/2$).  The quantum fluctuations between these four states 
[$S^{z}=\pm 1/2=\{\uparrow,\downarrow \}$ and $T^{z}=\pm 1/2=
\{\Uparrow,\Downarrow \}$] lead  to an unusual strongly correlated Fermi
liquid state in which the (real)-spin and pseudo-spin are totally entangled.
~\cite{Zar03}
In contrast to common spin Kondo physics observed in QD's, 
this new state possesses a higher symmetry, SU(4), corresponding to the 
total internal degrees of freedom of the double QD: $\{\uparrow \Uparrow, 
\downarrow \Uparrow, \uparrow \Downarrow,\downarrow \Downarrow \}$.  
The screened magnitude is now the {\em hyperspin} 
$\hat{M}\equiv\sum_{a,b} (\hat{S}^a+1/2)(\hat{T}^b+1/2)$.  Importantly, the
associated Kondo temperature $T_K^{\rm SU(4)}$ is much {\em higher} than in
the common spin-1/2 Kondo effect in a QD, which makes the observation
 of this spin and pseudo-spin entangled state more accessible.~\cite{Zar03} 
Strong entanglement of charge and spin flip events is also possible in a
single-electron box (metallic grain) coupled to a lead via a smaller
quantum dot in the Kondo regime.~\cite{Karyn03a,Karyn03b} Here, the spin Kondo
physics stems from the screening of the spin of the small dot while the
pseudo-spin Kondo physics emerges when charging states of the grain with
(charge) $Q=0$ and $Q=e$ are almost degenerate.

The most prominent feature of the Kondo effect 
is the phase coherence experienced by the electrons that 
participate in the many-body correlated state. Therefore, it 
is thus of great interest to have access to the phase of the 
transmission amplitude in order to give a fully characterization of
the transport properties. The widely known Aharonov-Bohm (AB) 
effect\cite{aha59} provides us a valuable tool to investigate 
quantum coherence of electrons.  When the coherence of a 
circulating electron wave packet enclosing a magnetic flux 
$\Phi$ is preserved, the result is an extra flux-dependent phase 
shift ($\phi$). In the simplest realization of an AB interferometer, 
an incoming electronic wavefunction splits into two paths, which join 
again into the outgoing electronic wavefunction. Applying a magnetic 
flux which threads this closed geometry, the outgoing wavefunction 
acquires a flux-dependent phase, $\phi=2\pi\Phi/\Phi_0$, where $\Phi=B/S$ 
is the flux, $B$ is the applied magnetic field, $S$ is the enclosing surface, 
and $\Phi_0=h/e$ the flux quantum. As a consequence, 
the transmission is a periodic function of $\phi$. 

\begin{figure}
\centering
\includegraphics*[width=85mm]{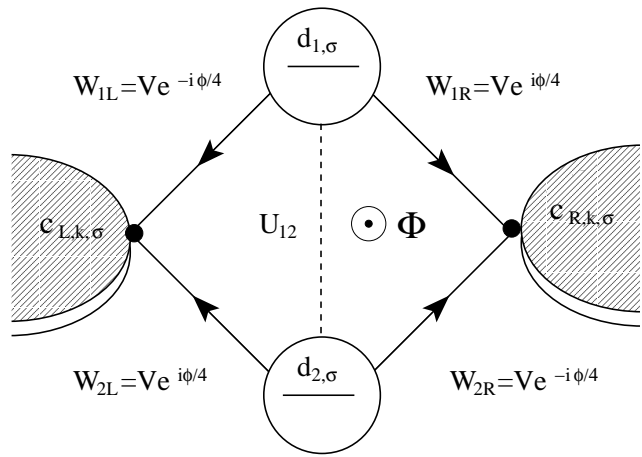}
\caption{ Sketch of the Aharonov-Bohm interferometer
containing two quantum dots attached to two leads.
The arrowed straight line indicates
the tunnel coupling and the dashed line represents the interdot
Coulomb interaction.}
\label{scheme}
\end{figure}

In this work, we consider a double quantum dot embedded in a prototypical
mesoscopic interferometer threading a magnetic flux
$\Phi$, see Fig.~\ref{scheme}.~\cite{Holl01}
Our motivation to investigate this system is twofold: 
(i) there are striking effects, such as Fano resonances,
which arise already in the noninteracting 
case~\cite{sha94,kub02,kan04} and, more interestingly, (ii) 
as the interdot interaction gets stronger,
the local density of states  on the double QD changes 
drastically.~\cite{boe02} {\em Here, we provide 
a unified picture of the combined
influence of wave interference, Kondo effect, and interdot
interaction on the electronic transport through a double QD
in and out of equilibrium.}

As we anticipated, the physical scenario in our setup will
depend much on the strength of the capacitive 
interdot coupling between the two dots. 
When the interdot Coulomb energy is negligible
each QD can accommodate one electron and both spins become screened. 
We find that each QD develops a Kondo resonance at the 
Fermi level $E_F=0$. Their interference causes a very 
narrow {\em dip} in the differential conductance $\mathscr{G}\equiv dI/dV_{\rm dc}$ 
except at $\phi\approx 0$ (mod $2\pi$). In the limit of strong interdot Coulomb
interaction there are two degenerate charge states described
by the pseudospin $T^{z}=\pm 1/2$. The presence of flux allow us to explore 
two interesting situations, (i) when the pseudo-spin is a good quantum number
[$\phi=\pi$ (mod $2\pi$)] and (ii) when the pseudo-spin is not conserved
during tunneling. In the former case the SU(4) Kondo state is fully developed
whereas far away from this symmetry point the conventional SU(2) Kondo physics
arises. In addition, we will show that $\mathscr{G}$ shows a zero bias anomaly
(ZBA) instead of a dip when the interdot Coulomb energy is large, which is
suppressed as $\phi$ enhances and eventually disappears at the destructive
interference condition [when $\phi=\pi$ (mod $2\pi$)] resulting a complete
suppression of the tunneling current. Nevertheless, this fact does not prevent
to us to observe the highly symmetric SU(4) Kondo state since it survives even
away from $\phi=\pi$ (mod $2\pi$) where the differential conductance is not
totally suppressed.

This work is organized as follows: we begin in Sec.~\ref{model}
presenting the theory to treat both limits for the interdot
Coulomb interaction using different theoretical techniques.  
We derive the transport properties as well. In Sec.~\ref{results} we 
present our numerical results and their interpretation.  
Finally, we shall end up by summarizing our main 
conclusions in Sec.~\ref{conclusions}. 

\section{Theory}
\label{model}
The system that we consider is depicted in Fig.~\ref{scheme}.  It is a
closed-geometry AB interferometer, where electrons emitted from the leads
are never lost in surrounding gates.  Electrons traveling through the
device have to go either through the upper dot or through the lower dot
before being transmitted into either the left or the right electrode.
The enclosing area by the two paths is penetrated by a flux $\Phi$.  The
two reservoirs are Fermi seas of electrons described by
the Hamiltonian
\begin{equation}
\label{kondo-su4::eq:H0}
\varH_0 = \sum_{\ell=L,R}\sum_{k,\sigma}
\veps_{\ell,k} c_{\ell,k,\sigma}^\dag c_{\ell,k,\sigma} \,,
\end{equation}
where $c_{L(R),k,\sigma}^\dagger$ ($c_{L(R),k,\sigma}$) is the
creation (annihilation) operator for an electron in the state $k$ with
spin $\sigma$ in the lead $L(R)$.  The isolated dots are described by
$\varH_{\rm D}$:
\begin{equation}
\label{kondo-su4::eq:HD}
\varH_{\rm {D}}
= \sum_{i=1,2}\left[
  \sum_{\sigma}\varepsilon_{i}\,
  d^\dagger_{i,\sigma} d_{i,\sigma}
  + U_in_{i,\up}n_{i,\down}
\right]
+ U_{12}n_{1}n_{2}\,.
\end{equation}
The operator $d_{i,\sigma}^\dagger$ ($d_{i,\sigma}$) is the creation
(annihilation) operator, $\varepsilon_{i}$ is the level position,
$U_{i}$ are the \emph{intradot} Coulomb interaction, and
$n_{i,\sigma}=d_{i,\sigma}^\dag d_{i,\sigma}$ is the
occupation number on the dot $i$.
$U_{12}$ denotes the \emph{interdot} Coulomb interaction
between the dots. The tunneling between the dots and the leads is 
modeled by $\varH_{T}$:
\begin{equation}
\label{kondo-su4::eq:HT}
\varH_T = \sum_{j=1,2}\sum_{\ell=L,R}\sum_{k,\sigma}
W_{\ell,j}\, c_{\ell,k,\sigma}^\dag d_{j,\sigma} + h.c \,.
\end{equation}
The tunneling amplitude $W_{\ell,j}$ in
Eq.~(\ref{kondo-su4::eq:HT}) from the dot $j$ to the lead $\ell$ is
modulated by the external flux $\Phi$ threading the loop
(Fig.~\ref{scheme}) and given by
\begin{equation}
\label{kondo-su4::eq:W}
\begin{split}
& W_{L,1} = V_{L,1}e^{-i\phi/4} \,,\quad
W_{L,2} = V_{L,2}e^{+i\phi/4} \,,\\
& W_{R,1} = V_{R,1}e^{+i\phi/4} \,,\quad
W_{R,2} = V_{R,2}e^{-i\phi/4} \,,
\end{split}
\end{equation}
where $V_{\ell,j}$ is the amplitude in the absence of the flux and
$\phi\equiv2\pi\Phi/\Phi_{0}$ with $\Phi_{0}$ being the flux quantum
($\Phi_{0}=h/e$).
Then the total Hamiltonian is
$\varH_\mathrm{total}=\varH_0+\varH_D+\varH_T$.

To make the physical interpretations of our results more clear, we perform a
few simplifications. First of all, we assume identical dots and
symmetric junctions; i.e., $\veps_1=\veps_2\equiv\veps_d$,
$U_1=U_2\equiv{U}$, and $V_{L,1}=V_{L,2}=V_{R,1}=V_{R,2}\equiv{V}$.
This is only for the sake of simplicity.\cite{robust}
Furthermore, we consider the wide-band limit, in which the couplings are
independent of energy. Then, the hybridization of the dot levels with
the conduction band is well characterized by the parameters
\begin{equation}
\label{kondo-su4::eq:Gamma}
\Gamma_{\ell;i,j}(\phi) = \pi\rho_\ell\, W_{\ell,i}W_{\ell,j}^* \,,
\end{equation}
or, in the matrix notation,
\begin{equation}
\label{kondo-su4::eq:GammaMatrix}
\widehat\Gamma_L(\phi) = \Gamma_L
\begin{bmatrix}
1 & e^{-i\phi/2} \\
e^{+i\phi/2} & 1
\end{bmatrix} \,,\quad
\widehat\Gamma_R = \widehat\Gamma_L^* \,,
\end{equation}
where
\begin{math}
\Gamma_\ell \equiv \pi\rho_\ell V^2
\end{math}
with $\rho_\ell$ being the DOS in the lead $\ell$ at the
Fermi energy ($\rho_L=\rho_R=\rho_0$).

Since we are interested in Kondo correlations~\cite{rkky}, we shall mainly concentrate 
on the Kondo regime [intradot charging energy $U\to\infty$ and localized level
$-\veps_d\gg(\Gamma_L+\Gamma_R)$] for which the fluctuations of the charges in the
single dots are highly suppressed. For the interdot Coulomb interaction $U_{12}$, 
we will investigate two opposite limits, namely, (i) $U_{12}=0$ and (ii) $U_{12}=\infty$.  
In the former case, each dot is singly occupied ($\avg{n_1}=\avg{n_2}\approx 1$) and
behaves as separate magnetic (Kondo) impurities. In the latter case, 
the double quantum dot system contains just one electron ($\avg{n_1+n_2}=1$).  
These two limits induce striking differences between the resulting Kondo effects. 
Moreover, the interference modulated by the external flux $\phi$ threading 
the AB geometry leads to an even richer variation of the Kondo effects in either case. 
Our goal is to investigate thoroughly these scenarios. For this purpose we
employ different techniques: scaling analysis (valid for $T\gg T_K$), the
slave-boson mean-field theory (SBMFT, for $T\ll T_K$), and the numerical
renormalization group (NRG) method. We elaborate below on these approaches.

\subsection{Scaling analysis}

We derive effective Hamiltonians in the Kondo regime 
for the two limiting cases ($U_{12}=0$ and $U_{12}=\infty$) 
and discuss their qualitative features at equilibrium 
by means of the scaling theory. 
\subsubsection{Case $U_{12}\to 0$}
Firstly, we discuss the large capacitance limit between the
two dots ($U_{12}\rightarrow 0$). As we mentioned, when 
$U_{12}$ is vanishingly small (and yet $U_1,U_2\to\infty$), 
the two dots are both singly occupied: $\avg{n_1} = 
\avg{n_2} \approx 1$ and each dot can thus be regarded as a 
magnetic impurity with spin 1/2. In this situation we notice 
that it is convenient and provides a more transparent picture 
of the system to perform the following canonical transformation
\begin{equation}
\label{kondo-su4::eq:CT0}
\begin{bmatrix}
c_{1,k,\sigma}\\
c_{2,k,\sigma}
\end{bmatrix} = \frac{1}{\sqrt{2}}
\begin{bmatrix}
e^{+i\pi/4} & e^{-i\pi/4} \\
e^{-i\pi/4} & e^{+i\pi/4}
\end{bmatrix}
\begin{bmatrix}
c_{L,k,\sigma} \\
c_{R,k,\sigma}
\end{bmatrix} \,.
\end{equation}
Under this transformation, the Hamiltonian for the leads 
Eq.~(\ref{kondo-su4::eq:H0}) is rewritten 
as follows
\begin{equation}
\label{kondo-su4::eq:H022}
\varH_0 = \sum_{\mu=1,2}\sum_{k,\sigma}\veps_k\,
c_{\mu,k,\sigma}^\dag c_{\mu,k,\sigma} \,,
\end{equation}
while the tunneling Hamiltonian Eq.~(\ref{kondo-su4::eq:HT}) reads 
\begin{equation}
\label{kondo-su4::eq:HT22}
\varH_T = \sum_{i=1,2}\sum_{\mu=1,2}\sum_{k,\sigma}
V_{\mu,i}c_{\mu,k,\sigma}^\dag d_{i,\sigma} + h.c. \,,
\end{equation}
where
\begin{equation}
\begin{split}
& V_{1,1} = V_{2,2} = \cos\frac{\phi-\pi}{4} \,,\\
& V_{1,2} = V_{2,1} = \cos\frac{\phi+\pi}{4} \,;
\end{split}
\end{equation}
Now, the Schrieffer-Wolff transformation~\cite{Schrieffer66a} 
of the Hamiltonians Eqs.~(\ref{kondo-su4::eq:H022}), 
(\ref{kondo-su4::eq:HT22}), and (\ref{kondo-su4::eq:HD}) 
leads to the Kondo-like Hamiltonian
\begin{widetext}
\begin{multline}
\label{kondo-su4::eq:HKondo:11}
\varH_\Kondo = \varH_0
+ \frac{1}{4} J_1\left(\bfS_1 + \bfS_2\right)\cdot
\left[\psi_1^\dag(0)\bfsigma\psi_1(0)
  + \psi_2^\dag(0)\bfsigma\psi_2(0)\right]
\\\mbox{}
+ \frac{1}{4} J_2\left(\bfS_1 + \bfS_2\right)\cdot
\left[\psi_1^\dag(0)\bfsigma\psi_2(0)
  + \psi_2^\dag(0)\bfsigma\psi_1(0)\right]
\\\mbox{}
+ \frac{1}{4} J_3\left(\bfS_1 - \bfS_2\right)\cdot
\left[\psi_1^\dag(0)\bfsigma\psi_1(0)
  - \psi_2^\dag(0)\bfsigma\psi_2(0)\right]
- \frac{1}{4} J_4\bfS_1\cdot\bfS_2 \,.
\end{multline}
\end{widetext}
In Eq.~(\ref{kondo-su4::eq:HKondo:11}), we have adopted the
spinor representations
\begin{equation}
\psi_j = 
\begin{bmatrix}
d_{j,\up}\\ d_{j,\down}
\end{bmatrix} \,,\quad
\psi_{\mu,k} = 
\begin{bmatrix}
c_{\mu,k,\up}\\ c_{\mu,k,\down}
\end{bmatrix}\,,
\end{equation}
($j=1,2$ and $\mu=1,2$) according to which the spin operator on
the dot $j$ is given by
\begin{equation}
\frac{\hbar}{2}\bfS_j = \frac{\hbar}{2}\psi_j^\dag\bfsigma\psi_j \,,
\end{equation}
where $\bfsigma$ denotes the three Pauli matrices.

The coupling constants $J_i$ ($i=1,\cdots,4$) in
Eq.~(\ref{kondo-su4::eq:HKondo:11}) are given initially 
(in the RG sense) by
\begin{multline}
\label{kondo-su4::eq:RG:10}
J_1 = 2\varN\frac{|V|^2}{|\veps_d|} \,,\quad
J_2 = J_1\cos(\phi/2) \,,\\\mbox{}
J_3 = J_1\sin(\phi/2) \,,
\end{multline}
where $\varN$ is the spin degeneracy. Under the renormalization group 
transformation~\cite{Anderson70a}, these coupling constants scale as
\begin{multline}
\label{kondo-su4::eq:RG:11}
\frac{dJ_1}{d\ell}  = \rho_0\left(J_1^2 + J_2^2 + J_3^2\right),\,
\frac{dJ_2}{d\ell}  = 2\rho_0J_1J_2\,,\\
\frac{dJ_3}{d\ell}  = 2\rho_0J_1J_3\,,
\end{multline}
where $\ell=-\log{D}$ indicates the renormalization steps ($D$ is the
bandwidth). $J_4$ is given
by~\cite{jayaprakash,Affleck95a,Jones}
\begin{equation}
J_4\approx 2\rho_0 J_1^2(0)\Upsilon(D)(1+\cos\phi) \,,
\end{equation}
where $\Upsilon(D)$ is order $1$.  $J_{4}$ corresponds to 
a ferromagnetic RKKY coupling between the spins in the dots.

Under the renormalization group
transformation all the system flows to the strong coupling fixed point with
the ratios $J_2/J_1$, $J_3/J_1$, and $J_4/J_1$ remaining constant.
In particular, the solution for the initial conditions
(\ref{kondo-su4::eq:RG:10}) satisfies the simple properties
\begin{equation}
\label{kondo-su4::eq:RG:12'}
\frac{J_2}{J_1} = \cos(\phi/2) \,,\quad
\frac{J_3}{J_1} = \sin(\phi/2) \,,\quad
\end{equation}
with $J_1\to\infty$ according to the equation
\begin{equation}
\label{kondo-su4::eq:RG:11'}
\frac{dJ_1}{d\ell} = 2\rho_0J_1^2 \,.
\end{equation}
From Eq.~(\ref{kondo-su4::eq:RG:12'}) one can easily see that 
system behaves in distinctive ways for different values 
of flux $\phi$, especially, for $\phi=0$ (mod $2\pi$) and
$\phi=\pi$ (mod $2\pi$). In the absence of the external flux ($\phi=0$), 
$J_3=0$ while $J_1 = J_2 = J$ and $J_{4}=I=\rho_{0}J_{1}^2/2$.  Thus the
Kondo-like Hamiltonian (\ref{kondo-su4::eq:HKondo:11}) is reduced to
\begin{widetext}
\begin{equation}
\label{kondo-su4::eq:HKondo:13}
\varH_\Kondo = \varH_0
+ \frac{1}{4} J\left(\bfS_1 + \bfS_2\right)\cdot
\left[\psi_1(0)+\psi_2(0)\right]^\dag\bfsigma
\left[\psi_1(0)+\psi_2(0)\right]
- I \bfS_1\cdot\bfS_2.
\end{equation}
\end{widetext}
This is the two-impurity (characterized by the two spins
$\frac{\hbar}{2}\bfS_1$ and $\frac{\hbar}{2}\bfS_2$) Kondo model coupled
to a single conduction band (characterized by $\psi_1+\psi_2$ or
equivalently $c_{1,k,\sigma}+c_{2,k,\sigma}$). The two spins are coupled to
each other ferromagnetically (-I $\bfS_1\cdot\bfS_2$, with $I>0$). 
Due to the ferromagnetic coupling and to the fact that both spins 
are coupled to the same conduction band, the total spin is 
underscreened at $T\to 0$.~\cite{jayaprakash} Note that a strong RKKY 
interaction may arise from our peculiar geometry since both QD's are directly 
connected to a single channel in the leads. Nevertheless, in an actual 
experimental situation~\cite{hol03,Holl01} the QD's are far apart 
and the RKKY interaction may be negligible. Furthermore, 
slightly above $\phi=0$  (mod $2\pi$), even for a \emph{large} 
ferromagnetic coupling $|I|\gg T_K=D\exp{(-1/2\rho_0 J)}$,
the spins of the dots added in a $S=1$ state become effectively 
screened.~\cite{jayaprakash}  

For the flux $\phi=\pi$ (mod $2\pi$), the coupling constant $J_2=0$  while
$J_1=J_3=J_4/2\equiv J$.  Then the Kondo-like Hamiltonian
\eqref{kondo-su4::eq:HKondo:11} is reduced to
\begin{multline}
\label{kondo-su4::eq:HKondo:14}
\varH_\Kondo = \varH_0
+ \half J\bfS_1\cdot\psi_1^\dag(0)\bfsigma\psi_1(0) \\\mbox{}
+ \half J\bfS_2\cdot\psi_2^\dag(0)\bfsigma\psi_2(0)
- \half J\bfS_1\cdot\bfS_2 \,.
\end{multline}
This model is clearly distinguished from the one in the previous case of
$\phi=0$ (mod $2\pi$) cf. Eq.~\eqref{kondo-su4::eq:HKondo:13}.  
The two impurity spins, $\frac{\hbar}{2}\bfS_1$ and 
$\frac{\hbar}{2}\bfS_2$, of magnitude $1/2$ 
are coupled to two independent conduction bands, $\psi_1$ and $\psi_2$ (or
equivalently $c_{1,k,\sigma}$ and $c_{2,k,\sigma}$), individually.  The
ferromagnetic coupling in Eq.~\eqref{kondo-su4::eq:HKondo:14} does not
play any significant role in this case, because its coupling strength
($I$) is the same as the exchange coupling between the localized spins
and the itinerant spins.  Therefore, the model
\eqref{kondo-su4::eq:HKondo:14} corresponds to the usual single-channel
spin 1/2 Kondo model.

The coupling constants scales according to the renormalization group
equation
\begin{equation}
\label{kondo-su4::eq:RG:14}
\frac{dJ}{d\ell} = 2\rho_0 J^2\,,
\end{equation}
and the Kondo temperature is given by
\begin{equation}\label{eq_tksu2}
T_K \sim D\exp\left(-\frac{1}{2\rho_0J}\right) \,.
\end{equation}
In the general case ($\phi\neq 0, \pi$), the two localized spins
$\frac{\hbar}{2}\bfS_1$ and $\frac{\hbar}{2}\bfS_2$ are coupled to two
conduction bands $\psi_1$ and $\psi_2$, let alone the ferromagnetic
coupling with each other.  Unlike the previous, special case
of $\phi=\pi$, the two conduction bands are not independent any longer;
see Eq.~\eqref{kondo-su4::eq:HKondo:11}.  This fact makes 
the physical interpretation of the model rather involved. However, 
the renormalization group flow  [see Eqs.~\eqref{kondo-su4::eq:RG:12'} and
\eqref{kondo-su4::eq:RG:11'}] and the results from the numerical
renormalization group method (see below)
 suggest that \emph{for any finite flux ($\phi\neq 0$), 
the two localized spins are fully screened out at zero temperature.}

\subsubsection{Case $U_{12}\to \infty$}

We now investigate the limit of $U_{12}\to\infty$ where the system 
properties change completely. In this case, only one electron is 
acommodated in the whole  double QD system, i.e., $\avg{n_1 + n_2} 
\approx 1$ having either  spin $\up$ or spin $\down$. The orbital 
degrees of freedom (pseudo-spin) play as significant a role 
as the spin, and the double QD behaves as an impurity with four 
degenerate levels with different tunneling amplitudes depending 
on the applied flux. Due to the orbital degrees of freedom involved 
in the interference, the symmetry of the wavefunction is crucial.  
Therefore, in this limit, it is more useful to work with a representation 
in terms of the symmetric (even) and antisymmetric (odd) combinations 
of the localized and delocalized orbital channels.~\cite{izu}

In accordance with these observations, we take the following canonical
transformations:
\begin{equation}
\label{kondo-su4::eq:CT1}
\begin{bmatrix}
d_{e,\sigma}\\ id_{o,\sigma}
\end{bmatrix} = \frac{1}{\sqrt 2}
\begin{bmatrix}
1 & 1\\ 1 & -1
\end{bmatrix}
\begin{bmatrix}
d_{1,\sigma}\\ d_{2,\sigma}
\end{bmatrix} \,,
\end{equation}
for the QD electrons, and
\begin{equation}
\label{kondo-su4::eq:CT2}
\begin{bmatrix}
c_{e,k,\sigma}\\ c_{o,k,\sigma}
\end{bmatrix} = \frac{1}{\sqrt 2}
\begin{bmatrix}
1 & 1\\ 1 & -1
\end{bmatrix}
\begin{bmatrix}
c_{L,k,\sigma}\\ c_{R,k,\sigma}
\end{bmatrix}\,,
\end{equation}
for the conduction electrons.

Then we identify the pseudo-spin up (down) as the electron occupying the
even (odd) orbital.  More explicitly, taking the four-spinor
representation
\begin{equation}
\label{kondo-su4::eq:spinor}
\psi_d^\dagger = 
\begin{bmatrix}
d_{e,\up},d_{e,\down},
d_{o,\up}, d_{o,\down}
\end{bmatrix},
\end{equation}
the spin and orbital pseudo-spin operators are given by
\begin{equation}
\frac{\hbar}{2}\bfS = \frac{\hbar}{2}\psi_d^\dag\bfsigma\psi_d \,,\quad
\frac{\hbar}{2}\bfT = \frac{\hbar}{2}\psi_d^\dag\bftau\psi_d \,,
\end{equation}
respectively, where $\bfsigma$ ($\bftau$) are Pauli matrices operating
on the spin (pseudo-spin) space. Notice that in this even/odd basis the dot
pseudo-spin has been rotated: $\overline{T}^{x}\rightarrow T^{z}$, 
$\overline{T}^{y}\rightarrow -T^{y}$ and $\overline{T}^{z}\rightarrow T^{x}$ 
whereas the spin remains invariant.

In terms of the new operators $d_{e,\sigma}$, $d_{o,\sigma}$,
$c_{e,k,\sigma}$ and $c_{o,k,\sigma}$, the total Hamiltonian, 
$\varH_{\rm total}$ is rewritten as follows
\begin{widetext}
\begin{multline}
\label{kondo-su4::eq:HD:2}
\varH_D =\sum_{\alpha=e,o}\sum_{k,\sigma}
\veps_{k,\sigma}c_{\alpha,k,\sigma} +
\sum_{\alpha=e,o}\left[
  \sum_\sigma\eps_d\, d_{\alpha,\sigma}^\dag d_{\alpha,\sigma}
  + \half\left(U + U_{12}\right)n_{\alpha,\up}n_{\alpha,\down}
\right]+ \frac{1}{4}(U+3U_{12})n_e n_o \\\mbox{}\hfil%
- \frac{1}{4}(U-U_{12})\left(d_e^\dag\bfsigma d_e\right)\cdot
\left(d_o^\dag\bfsigma d_o\right)- \half(U-U_{12})
\left(d_{e,\up}^\dag d_{e,\down}^\dag d_{o,\down} d_{o,\up}
  + h.c.\right)+\sum_{\alpha=e,o}\sum_{k,s}
V_\alpha\, c_{\alpha,k,s}^\dag d_{\alpha,s}
+ h.c. \,, 
\end{multline}
\end{widetext}
where
\begin{equation}
\label{Model::eq:V}
V_e \equiv 2V\cos(\phi/4) \,,\quad
V_o \equiv 2V\sin(\phi/4) \,.
\end{equation}
Therefore, the even (odd) orbitals are coupled only to the
even(odd)-symmetric combinations of the conduction bands.

To examine the low-energy properties of the
system, we obtain for all values of $\phi$
the following effective Hamiltonian by 
performing a Schrieffer-Wolf transformation~\cite{Schrieffer66a}:
\begin{widetext}
\begin{multline}
\label{kondo-su4::eq:HKondo:1}
\varH_\Kondo = \varH_0
+ \frac{1}{4} J_1 \bfS\cdot(\psi^\dag\bfsigma\psi)
+  \frac{1}{4} J_2 \bfS\cdot(\psi^\dag\bfsigma\overline{\bftau}^\bot\psi)
\cdot\bfT^\bot
+ \frac{1}{4} J_1 \bfS\cdot(\psi^\dag\bfsigma\overline{\tau}^z\psi) \overline{T}^2
\\\mbox{}
+ \frac{1}{4} J_2 (\psi^\dag\overline{\bftau}^\bot\psi)\cdot\bfT^\bot
+ \frac{1}{4}  J_3 (\psi^\dag \overline{\tau}^z\psi)\overline{T}^z
+ \frac{1}{4} J_4 \left[
  \bfS\cdot(\psi^\dag\bfsigma\overline{\tau}^z\psi)
  + \bfS\cdot(\psi^\dag\bfsigma\psi)\overline{T}^z
\right]
- J_5\overline{T}^z \,,
\end{multline}
\end{widetext}
where $\varH_0$ is the first term in Eq.~(\ref{kondo-su4::eq:HD:2})
and $\psi=[\psi_{e\up},\psi_{e\down},\psi_{o\up},\psi_{o\down}]$
is the spinor of the itinerant electrons.
Here the effective coupling constants $J_i$ ($i=1,\dots,6$) are
initially (in the RG sense) given by
\begin{multline}
J_1 = J_3 = 2\varN\frac{|V|^2}{|\veps_d|} \,,\quad
J_2 = J_1\sin(\phi/2) \,,\\
J_4 = J_1\cos(\phi/2) \,,
\end{multline}
and scale according to the RG equations 
(up to the second order in $J$'s)
\begin{eqnarray}
\label{kondo-su4::eq:RG}
&&\frac{dJ_1}{d\ell}  = 2\rho_0\left(J_1^2 + J_2^2 + J_4^2\right),\,
\frac{dJ_2}{d\ell}  = \rho_0J_2(3J_1+J_3),\,\nonumber 
\\
&&\frac{dJ_3}{d\ell}  = 4\rho_0J_2^2 ,\,
\frac{dJ_4}{d\ell}  = 4\rho_0J_1J_4 \,.
\end{eqnarray}
$J_5$ is given
and by~\cite{Affleck95a,Jones,jayaprakash}
\begin{multline}
J_5 = 4\rho_0|V|^2\cos(\phi/2)\ln\frac{\veps_d+D}{\veps_d-D} \\\mbox{}%
+ 8[\rho_0J_1(\ell=0)]^2\Upsilon(D)\cos(\phi/2) \,.
\end{multline}

As one can see from the RG equations (\ref{kondo-su4::eq:RG}), in general,
each coupling constant in Eq.~(\ref{kondo-su4::eq:HKondo:1}) scales
differently under the renormalization group transformation 
for typical behaviors of the solutions at
different values of flux $\phi$.  Importantly, we show now that
the system exhibits a crossover from $0$-flux to $\pi$-flux.
Near the $0$-flux [$\phi\approx 0\pmod{2\pi}$],
the double QD odd orbital is completely decoupled from the odd-symmetric lead 
and only the even orbital is coupled to the even-symmetric 
conduction lead with $V_{e}=2V$ [see Eq.(\ref{Model::eq:V})].
Equation (\ref{kondo-su4::eq:HKondo:1}) then reduces to a
model involving only the spin in the even orbital 
$\frac{\hbar}{2}\bfS_e$ (not $\frac{\hbar}{2}\bfS$)
\begin{multline}
\label{kondo-su4::eq:HKondo:3}
\varH_\Kondo
= \varH_0
+ J\bfS_e\cdot(\psi_{e}^\dag\bfsigma\psi_{e})
(1+\overline{T}^z) \\\mbox{}
+ \frac{1}{4} J(\psi_{e}^\dag\psi_{e})\overline{T}^z
- J_5\overline{T}^z\,,
\end{multline}
where $J=2|V|^2/|\varepsilon_d|$. This model was already analyzed 
in Ref.~\onlinecite{zarand95}, where it was shown
that the ground state corresponds to a Fermi liquid state with a greatly 
enhanced Kondo temperature $T_K^{\rm SU(2)}=D\exp(-1/4\rho_0J)$ (due to a
coupling doubling of the even orbital to the even-symmetric conduction lead, 
i.e., $V_{e}=2V$) and the orbital pseudo-spin gets frozen completely,
$\overline{T}^z=1$. ($J_5$ does not flow to the strong coupling regime). 
One can easily see that the model (\ref{kondo-su4::eq:HKondo:3}) is 
equivalent to the twofold orbitally degenerate Anderson model described 
by the common \emph{SU(2) Kondo physics}.~\cite{boe02} 

Near the $\pi$-flux [$\phi\approx\pi\pmod{2\pi}$] the exchange couplings are
$J=J_1=J_2=J_3$, and $J_{4}=J_{5}=0$. The corresponding Kondo-like Hamiltonian reads
\begin{multline}\label{su4hamiltonian}
\varH_K
= \frac{J}{4}[\bfS\cdot(\psi^\dag\bfsigma\psi)+
(\psi^\dag\bftau\psi)\cdot\bf{T}
\\
+\bfS\cdot(\psi^\dag\bfsigma\bftau\psi)\cdot\bf{T}] \,.
\end{multline}
This is the celebrated \emph{SU(4) Kondo model}, where the spin and the orbital
degrees of freedom become entangled due to the third term in
 Eq.~(\ref{su4hamiltonian}).
The RG equation reads 
\begin{equation}
dJ/d\ell = 4\rho_0J^2\,,
\end{equation}
leading to
\begin{equation}\label{eq_tksu4}
T_K^{\rm SU(4)}=D\exp(-1/4\rho_0J)\,.
\end{equation}
As the flux departs from $\pi$, the degeneracy of the even and odd
orbitals is lifted and the SU(4) symmetry is broken, much like a single
Kondo impurity in the presence of a Zeeman splitting.~\cite{j5}
The crossover from the SU(4) to the SU(2) Kondo model occurs at a 
given critical flux $\phi_{c}$. From our NRG calculation (see below) 
we estimate $\phi_{c} \approx 0.75\pi$.
 
This discussion demonstrates the existence of high-symmetry Kondo
states in double quantum systems with interdot interaction
in the presence of an Aharonov-Bohm flux. We have shown that the magnetic flux
critically alters the properties of the ground state, resulting in a smooth
transition from SU(2) to SU(4) Kondo physics. Below, we prove that
the {\em differential conductance} would indicate the principal
features of this effect. This is important since it would serve as
a means of experimental detection.
 
\subsection{Slave-boson mean-field theory}

In this section, we adopt the so-called slave-boson mean-field theory
which captures the main physics of the Kondo problem~\cite{sbmft} at 
sufficiently low temperatures ($T\ll T_K$).  The SBMFT corresponds to 
the leading order in a $\varN$-large expansion, where $\varN$ is 
the degeneracy of each site. Such a SBMFT has been
recently applied to study the Kondo effect in nonequilibrium
situations~\cite{eto01,dong02,avish03} and in double QD's 
systems~\cite{geo99,ram00,ros02}.

\subsubsection{Case $U_{12}\to 0$}

Firstly we consider the case of vanishing interdot Coulomb interaction $U_{12}=0$.
We express the two-impurity Anderson model ($\varH_{\rm total}$)
in terms of the slave boson operators. This way the fermionic 
operator of each dot is written as a combination of a 
pseudofermion and a boson operator: $d_{i,\sigma}=b^\dagger_i f_{i,\sigma}$,
where $f_{i,\sigma}$ is the pseudofermion which annihilates one 
``occupied state'' in the $i$th dot and $b^\dagger_i$ is a boson 
operator which creates an ``empty state'' in the $i$th dot.
We include two constraints to prevent double occupation in each QD in the
limit $U_{1},U_{2}\to\infty$ by using two Lagrange multipliers $\lambda_{1},\lambda_{2}$. 
Thus, the Hamiltonian in the slave boson language reads:

\begin{widetext}
\begin{multline}
\label{hamiltonian1}
\varH_\mathrm{SB}
= \varH_0 + \sum_{i=1,2}\sum_{\sigma}\veps_{i,\sigma}
f_{i,\sigma}^\dag f_{i,\sigma}
+ \frac{1}{\sqrt{\varN}}\sum_{i=1,2}\sum_{\ell=L,R}\sum_{k,\sigma}
\left(\overline{W}_{\ell,i}\,
  c_{\ell,k,\sigma}^\dag b_i^\dag f_{i,\sigma} + h.c.\right)
\\\mbox{}
+ \sum_{i=1,2}\lambda_i\left(\sum_\sigma f_{i,\sigma}^\dag f_{i,\sigma}
  + b_i^\dag b_i - 1\right) \,,
\end{multline}
\end{widetext}
where $\overline{W}_{\ell,i}={W}_{\ell,i} \sqrt{\varN}$.
The hallmark of the SBMFT consists of replacing the boson operator 
by its classical (nonfluctuating) average.
\begin{math}
b_i(t)/\sqrt{\varN} \rightarrow
\langle b_i \rangle/\sqrt{\varN}
\equiv \tilde{b}_i
\end{math}
thereby neglecting
charge fluctuations in each dot. This approximation is exact in the
limit $\varN\rightarrow\infty$, and it corresponds to $\mathscr {O}(1)$ in a
$1/\varN$ expansion. At zero temperature $T=0$, it correctly describes
spin fluctuations (Kondo regime). Then, the mean field Hamiltonian
is given by
\begin{multline}
\varH_\mathrm{MF}= \varH_0
+ \sum_i\sum_{\sigma}\tilde\veps_{i,\sigma}
f_{i,\sigma}^\dag f_{i,\sigma}
\\\mbox{}
+ \sum_i\sum_{\ell,k,\sigma}
\left(\widetilde{W}_{\ell,i}\,
  c_{\ell,k,\sigma}^\dag f_{i,\sigma} + h.c.\right)
\\\mbox{}
+ \sum_i\lambda_i\left(\varN|\tilde{b}_i|^2-1\right) \,,
\end{multline}
where $\widetilde{W}_{\ell,i}=\tilde{b}_i\overline{W}_{\ell,i}$.
We obtain a quadratic Hamiltonian containing
four parameters ($\tilde{b}_{1,2}$ and renormalized levels  
$\tilde{\varepsilon}_{1,2}=
\varepsilon_{d}+\lambda_{1,2}$) to be determined 
from mean field equations.~\cite{sbmft,ram00} These mean-field 
equations are the constraints for the dot
$i=1,2$:
\begin{equation}
\label{kondo-su4::eq:premean1}
\sum_{\sigma}
\langle f^\dagger_{i,\sigma}(t)f_{i,\sigma}(t)\rangle
+ \varN|\tilde{b}_i|^2 = 1\,,
\end{equation}
and the equations of motion (EOM) of the boson fields:
\begin{equation}
\label{kondo-su4::eq:premean2}
\sum_{\ell,k,\sigma}\widetilde{W}_{\ell,i}
\left\langle c_{k,\sigma}^\dagger(t)f_{i,\sigma}(t)\right\rangle
+ \lambda_i \varN|\tilde{b}_{i}|^2 = 0 \,.
\end{equation}

The next step is to write these mean field equations in terms of
nonequilibrium Green functions. The lesser dot-dot Green function 
is ($i\in 1,2$) $G^<_{i,\sigma}(t-t')=-i\langle f^{\dagger}_{i,\sigma}(t')
 f_{i,\sigma}(t)\rangle$, and the corresponding lesser lead-dot Green 
function is $G^<_{i,\sigma;\ell,k,\sigma}(t-t')=-i\langle 
c^\dagger_{\ell,k,\sigma}(t') f_{i,\sigma}(t)\rangle$.
 By applying the equation-of-motion (EOM) technique and the 
analytical continuation rules~\cite{lan76,mei93} we can relate 
the lesser lead-dot Green function with the dot-dot 
Green function. Eventually, the explicit form of the Green's functions can 
be found easily using the EOM technique. This way we close the set 
of mean field equations, which are self-consistently 
solved for each set of parameters (the dot levels $\veps_i$, the tunneling 
amplitudes $V_{\ell,i}$, the flux $\phi$, the band width $D$, and the 
applied dc bias $V_\mathrm{dc}$).

At zero bias we can derive analytical expressions 
of $T_K$ ($\phi$-dependent) within the SBMFT. 
E.g., for $\pi$-flux we get $T_K^{\rm SU(2)}=
D \exp(-\pi |\varepsilon_d|/2\Gamma)$ 
($\Gamma=\pi\rho_{0}|V|^2$ is the hybridization width).
As expected, it is in agreement with scaling theory,
see Eq.~(\ref{eq_tksu2}).

\subsubsection{Case $U_{12}\to\infty$}

For $U_{12}\to\infty$ only one dot can be charged at a given time. 
In this case we introduce one boson field and one constraint that preserves 
the condition $\avg{n_{1}+n_{2}}=1$. The rest of the calculation
follows the lines exposed above. We find for the Kondo temperature at $\phi=\pi$
$T_K^{\rm SU(4)}=(D/\sqrt{2}) \exp(-\pi |\varepsilon_d|/4\Gamma)$
[cf. Eq.~(\ref{eq_tksu4})].

\subsubsection{Transport properties}

In the reminder, we describe how to calculate the current 
through the double QD system  within the SBMFT.  
The simplicity of our approach allows us to write 
the current using the Landauer-B\"uttiker formula:
\begin{eqnarray}\label{current}
I= \frac{2e}{\hbar}\int
\frac{d\varepsilon}{2\pi} \mathscr{T}(\varepsilon,V_{\rm
dc})\left[f_{L}(\varepsilon)-f_{R}(\varepsilon)\right],
\end{eqnarray}
where $\mathscr{T}(\varepsilon,V_{\rm dc})$ is the transmission
probability which depends on renormalized parameters. Following Meir
and Wingreen~\cite{mei92}, the transmission through this system can be
obtained using 
\begin{equation}
\mathscr{T} = \mathrm{Tr}\{ \widehat{G}^{a}\widetilde\Gamma_{R}
\widehat{G}^{r}\widetilde\Gamma_{L} \} \,.
\end{equation}
Here $\widehat{G}^{a(r)}$ is the matrix of the
advanced (retarded) Green's function for the dot electrons; i.e.,
\begin{math}
G_{i,j;\sigma}^{r/a}(t) = \mp i\,\theta(\pm t)
\langle\{d_{i,\sigma}(t),d_{j,\sigma}^\dag(0)\}\rangle
\end{math}.
$\widetilde\Gamma_\ell$ is the matrix of the renormalized hybridization
parameters; i.e., 
\begin{math}
\widetilde\Gamma_{\ell;i,j}=\pi\rho_\ell
\overline{W}_{\ell,i}\overline{W}_{\ell,j}^* \tilde{b}_i\tilde{b}_j^*
\end{math} for $U_{12}=0$ and
\begin{math}
\widetilde\Gamma_{\ell;i,j}=\pi\rho_\ell
\overline{W}_{\ell,i}\overline{W}_{\ell,j}^* |\tilde{b}|^2
\end{math} for $U_{12}=\infty$.

For identical dots and symmetric junctions the transmission 
probability is given by
\begin{equation}
\label{eq_tprob}
\mathscr{T}(\epsilon) = \frac{
  \widetilde\Gamma^2 \left[ (\epsilon - \tilde\veps_d)^2
      \cos^2 \frac{\phi}{2} \right] }{
    \left[ (\epsilon - \tilde\veps_d)^2
      - \left(\frac{\widetilde\Gamma}{2}\right)^2
      \sin^2 {\phi \over 2} \right]^2 + (\epsilon - \tilde\veps_d)^2
    \widetilde{\Gamma}^2
  }\,,
\end{equation}
regardless whether  $U_{12}=0$ or $U_{12}=\infty$.  Of course, the
renormalized coupling $\widetilde{\Gamma}$ in the above equation
should be obtained according to the different set of mean-field 
equations, depending on whether $U_{12}=0$ or $U_{12}=\infty$. We notice that
Eq.~(\ref{eq_tprob}) was previously obtained in Refs.~\cite{sha94,kub02} for the 
noninteracting case. 

The expression for the nonlinear conductance is straightforward from the
current expression $I$: $\mathscr{G}=dI/dV_{\rm dc}$. In the same way, the
linear conductance $\mathscr{G}_{0}$ is determined upon insertion of the total
transmission evaluated at the Fermi energy into the well known formula
\begin{equation}
\mathscr{G}_{0}=\frac{2e^2}{h}\mathscr{T}(E_F) \,.
\end{equation}

\subsection{Numerical Renormalization Group}

SBMFT does not take fully into account real charge fluctuation effects.
In order to confirm our previous results and make quantitative
predictions we also use the NRG
procedure~\cite{Wilson75a,Krishna-murthy80a,cos94,Hofstetter00a}.

Following the standard NRG
procedures~\cite{Wilson75a,Krishna-murthy80a,cos94}, we evaluate the
various physical quantities from the recursion relation ($N\geq 0$)
\begin{multline}
\label{sds-kondo::eq:HNRG}
\vtilH_{N+1} = \sqrt\Lambda\, \vtilH_N  \\\mbox{} 
+ \xi_{N+1}\sum_{\mu=e,o}\sum_{\sigma=\up,\down}
\left(f_{\mu,N,\sigma}^\dag f_{\mu,N+1,\sigma} + h.c.\right)\,,
\end{multline}
with the initial Hamiltonian given by
\begin{equation}
\label{sds-kondo::eq:H0}
\vtilH_0 = \frac{1}{\sqrt\Lambda} \Biggl[ \vtilH_D +
\sum_{\mu=e,o}\sum_\sigma\tilV_\mu \left( d_\sigma^\dag f_{\mu,0,\sigma}
  + h.c. \right)
\Biggr] \,.
\end{equation}
Here the fermion operators $f_{\mu,N,\sigma}$ have been introduced as a
result of the logarithmic discretization and the accompanying canonical
transformation, $\Lambda$ is the logarithmic discretization parameter
(we choose $\Lambda=2$),
\begin{equation}
\xi_N \equiv
\frac{1-\Lambda^{-N}}{\sqrt{[1-\Lambda^{-(2N-1)}][1-\Lambda^{-(2N+1)}]}} \,,
\end{equation}
and
\begin{equation}
\vtilH_D \equiv \zeta \frac{\varH_D}{D}
\end{equation}
with $\zeta=\frac{2}{1+1/\Lambda}$.
The coupling constants $\tilV_e$ and $\tilV_o$, respectively, are given by
\begin{gather}
\tilV_e \equiv 4\zeta
\sqrt{\frac{2\Gamma}{\pi D}}\cos(\phi/4), \,\\
\tilV_o \equiv 4\zeta
\sqrt{\frac{2\Gamma}{\pi D}}\sin(\phi/4) \,.
\end{gather}
The Hamiltonians $\vtilH_N$ in Eq.~(\ref{sds-kondo::eq:HNRG}) have been
rescaled for numerical accuracy.  The original Hamiltonian is recovered
by
\begin{equation}
\frac{\varH}{D} = \lim_{N\to\infty}\frac{\vtilH_N}{\varS_N}\,,
\end{equation}
with
\begin{math}
\varS_N \equiv \zeta\Lambda^{(N-1)/2} \,.
\end{math}

In the following, we study the local Green's functions
(with which the linear conductance is calculated) and the dynamic spin
susceptibility.  To improve accuracy at higher energies, we adopt the
density-matrix NRG method (DM-NRG)~\cite{Hofstetter00a}.  In this
method, first usual NRG iterations are performed down to the energy
scale $\omega_N\equiv D\Lambda^{-N/2}\ll T_K$.  From the excitation
spectrum at this scale, the density matrix is constructed:
\begin{equation}
\rho = \sum_m e^{-E_m^N/\omega_N}\ket{m}_N\bra{m} \,,
\end{equation}
where $\ket{m}_N$ is the the eigenstate of $\varH_N$ with energy
$E_m^N$.  Then, the NRG iterations are performed again, but now at each
iteration $N'$, calculating the Green's function by
\begin{equation}
\label{kondo-su4-prb::eq:G1}
G_{\mu\sigma;\mu'\sigma'}(t)
= \frac{i}{\hbar}\theta(t)
\tr\rho_{N'}[d_{\mu,\sigma}(t),d_{\mu',\sigma'}^\dag] \,,
\end{equation}
where
\begin{equation}
\rho_{N'} \equiv \tr_{N>N'}\rho
\end{equation}
is the reduced density matrix for the cluster of size $N'$.  The Green's
function in Eq.~(\ref{kondo-su4-prb::eq:G1}) is valid at the frequency
scale $\omega\simeq\omega_{N'}$.  The spin susceptibility is calculated
in the same manner:
\begin{equation}
\chi(\omega) = -\frac{1}{\pi}\im\int_{-\infty}^\infty{dt}\;
e^{+i\omega t}\frac{1}{i\hbar}\tr\rho_{N'}\{s_z(t),s_z\} \,,
\end{equation}
where 
\begin{equation}
s_z\equiv\half\sum_\mu [d_{\mu,\up}^\dag d_{\mu,\up} -
d_{\mu,\down}^\dag d_{\mu,\down}] \,.
\end{equation}

\section{Numerical results}
\label{results}
We now present our results for the electronic transport in both limits
of the Coulomb interaction, $U_{12}\to 0$ and $U_{12}\to\infty$.  In the
numerical calculations, the model parameters are taken as follows:
symmetric couplings ($\Gamma_{1(2)L}=\Gamma_{1(2)R}=\Gamma/2$) and equal
level positions ($\varepsilon_{1}=\varepsilon_{2}=\veps_d$).
Throughout this paper, all the parameters are given in units of the bare
coupling $\Gamma$.  The energy cutoff is set as $D=60\Gamma$.

\begin{figure}\centering
\includegraphics*[width=85mm]{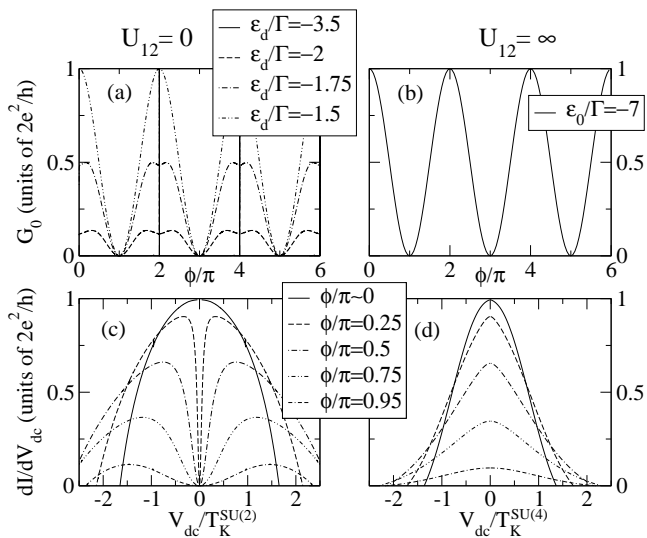}
\caption{\emph{SBMFT results:} Left panel ($U_{12}\to 0$): 
  (a) Linear conductance ($\mathscr{G}_{0}$) versus flux $\phi$
  for different level positions. When the Kondo state is formed (for
  $\veps_d=-3.5$) the $\mathscr{G}_{0}$ are delta-like peaks of
  height $1$ centered at even multiples of $\phi/\pi$.  (c) Curves for 
  $\mathscr{G}$ versus voltage bias for $\veps_d=-3.5$.
  Here we change the flux from 0 (full line) to $\pi$
  (dot-dot-dashed line). Right panel ($U_{12}\to\infty$): (b) linear and (d) 
  differential conductance. Energies are measured in units
  of $\Gamma=\pi\rho_{0}|V|^2=D/60$.}
\label{fig2}
\end{figure}

\subsection{Case $U_{12}\to 0$}
In the left panel of Fig.~\ref{fig2} we present our results when $U_{12}\to 0$ 
obtained with slave-boson mean-field theory.
First, we focus on the pure Kondo regime when 
$\varepsilon_d=-3.5$ and discuss both the linear conductance and the
nonlinear conductance (given by $\mathscr{G}_{0}\equiv\mathscr{G}(V_{\rm dc}=0)$
 and $\mathscr{G}\equiv dI/dV_{\rm dc}$, respectively).
The linear conductance [solid line in Fig.~\ref{fig2}(a)]
shows \emph{narrow} peaks due to constructive interference 
around $\phi\approx 0$ (mod $2\pi$) whereas transport is suppressed elsewhere. 
This is caused by the fact that the DOS of each dot has a resonance 
exactly at $E_F$. In the language of slave bosons 
this means  $\tilde{\varepsilon}_{1,2}=0$ and a SU(2) Kondo state is
 well formed. Therefore, these narrow peaks in $\mathscr{G}_{0}$ 
correspond to paths through the AB geometry with multiple 
windings around the enclosed flux. 
The width of each peak is given roughly by $\approx T_K$. 
Away from the constructive interference
condition, the transmission at the Fermi energy quickly vanishes.  
This unusual behavior is clarified with our calculations of the 
differential conductance. In Fig.~\ref{fig2}(c) we show the 
nonlinear conductance $dI/dV_{\rm dc}$ as a function of the bias voltage 
$V_{\rm dc}$ for different values of the flux and $\veps_d=-3.5$. 
In the absence of flux (or for even multiples of it) the
nonlinear conductance shows the usual zero-bias anomaly (ZBA), 
a narrow peak at $V_{\rm dc}=0$ that reaches the
unitary limit due to the constructive interference in the resonant
condition. Increasing $\phi$ does not affect the
Kondo resonance much so that the transmission probability $\mathscr{T}$ can be
written as a combination of a Breit-Wigner resonance for
$\tilde{\varepsilon}_d=0$ plus a Fano antiresonance.~\cite{sha94} A {\em dip}
at zero bias is then obtained [see Fig.~\ref{fig2}(c)].
~\cite{sha94,kan04} The width of this dip is $T_K(\phi)[1-\cos(2\phi)]$. It has
an oscillatory dependence on the applied flux. This result is in good 
agreement with the NRG calculations as shown in Fig.~\ref{fig3}(a). Here, we
plot $\mathscr{T}$ as a function of energy. 
It is worthwhile to note that $\mathscr{T}$ amounts to
$\mathscr{G}$ at low bias. 

For increasing $\varepsilon_d$ one enters 
the mixed-valence regime [see Fig.~\ref{fig2}(a)]. 
Although the results should be taken in a qualitative way, 
we find that the renormalized levels for
$\veps_d=-2,-1.75, -1.5$ are no longer at $E_{F}$ except when
$\phi\approx 0$ (mod $2\pi$). The transmission coefficient 
(and thereby the conductance) is extremely
sensitive to deviations of $\tilde\veps_d$ out of $E_F$. When
the bare level position is shifted toward the Fermi energy
 the renormalized levels for $\veps_d=-2,-1.75$ as 
a function of $\phi$  are not at $E_{F}$ 
except when $\phi\approx 0$ (mod $2\pi$) whereas for $\veps_d=-1.5$ 
they never reach $E_F$. In these cases due to the lack of a
 resonant condition at each dot, multiple windings
are less likely to occur and the conductance starts to resemble a
cosine-like function  generated by a combination of lower harmonics 
[see Fig.~\ref{fig2}(a), case $\veps_d=-1.5$].  
For $\veps_d=-2,-1.75$ we still
observe the sharp resonance at $\mathscr{G}(\phi\approx 0)$ (mod $2\pi$) due
to a quasiresonant condition when $\phi\approx 0$ (mod $2\pi$).
Finally,
for $\veps_d=-1.5$ the linear conductance has a trivial cosine dependence.

\subsection{Case $U_{12}\to\infty$}

Next, we elaborate on the numerical results for the limit of a strong
interdot Coulomb interaction $U_{12}\to\infty$ (right panel of Fig.~\ref{fig2}).
The results show that in this situation not only the spin 
fluctuates but also the pseudo-spin since just two charge 
states are allowed in the double QD system: $\{n_{1}=1, n_{2}=0\}$ and 
$\{n_{1}=0, n_{2}=1\}$. The fluctuations in both sectors
(spin and pseudo-spin) leads to the exotic SU(4) state close 
to $\phi=\pi$ (mod $2\pi$). 

We begin with the linear regime. Figure~\ref{fig2}(b) summarizes our
results for $\mathscr{G}_0$ as a function of the applied flux. We concentrate
on the pure Kondo regime and set $\veps_d=-7$ well below $E_F$.
Unlike the case of weak interdot Coulomb interaction 
[see Fig.~\ref{fig2}(a)], the linear conductance shows 
{\em broad} peaks at positions $\phi\approx 0$ (mod $2\pi$). 
In addition, the linear 
conductance only vanishes when the 
condition of destructive interference takes place. 
Let us investigate in in some detail the two limit cases 
$\phi\approx 0$ and $\phi\approx \pi$ (mod $2\pi$). 
In our RG analysis, we find for $\phi\approx 0$ (mod $2\pi$) 
that the ground state corresponds to the 
usual spin SU(2) Kondo effect with a greatly enhanced Kondo scale. 
Accordingly, the corresponding renormalized level
lies at $\tilde{\veps}_d=0$ leading to a shift of the scattering 
phase $\delta=\pi/2$. On the contrary, for $\phi=\pi$ (mod $2\pi$) we find
that the ground state is the highly symmetric SU(4) Kondo state with
 a renormalized level at $\tilde{\veps}_d\approx T_K^{\rm SU(4)}$,
 which implies $\delta=\pi/4$ to fulfill the Friedel-Langreth sum
 rule.~\cite{hew93} Quite generally, in a SU(N) problem the phase shift becomes
 $\delta=\pi/\varN$ in the limit of large $\varN$ and the Kondo resonance
 shifts up to $\tilde\veps_d\approx \pi\widetilde\Gamma/\varN$.~\cite{hew93} 
In the intermediate regime, when $0\lesssim\phi\lesssim\pi$,
the renormalized level takes on a positive value
$\tilde{\veps}_d<T_K^{\rm SU(4)}$. As a consequence, away 
from  $\phi\approx 0$ (mod $2\pi$) the resonant
condition is not satisfied (the renormalized level $\tilde{\veps}_d$ 
is not longer at $E_F$). In this situation electronic paths with 
multiple windings do not occur and the linear conductance consists 
of a cosine-like function [Fig.~\ref{fig2}(b)]. At finite $V_{\rm dc}$ 
the nonlinear conductance displays a ZBA which is quenched as $\phi$ 
decreases [see Fig.~\ref{fig2}(d)], unlike the dip found in the 
$U_{1 2}=0$ case. Eventually, for $\phi=\pi$ (mod $2\pi$) there is
no transport due to completely destructive interference.

We can compare our results shown in Fig.~\ref{fig2}(d) with those
obtained from NRG plotted in Fig.~\ref{fig3}(b).  Here, one can see that
$\mathscr{T}$ decreases as $\phi$ increases, which is consistent with
the results of the SBMFT.  Nevertheless, SBMFT overestimates the
decreasing rate of the ZBA. The NRG results show that while the peak
does not change appreciably for $\phi<\phi_{c}$, it decreases very
rapidly for $\phi>\phi_{c}$.

\subsection{Crossover}

\begin{figure}\centering
\includegraphics*[width=85mm]{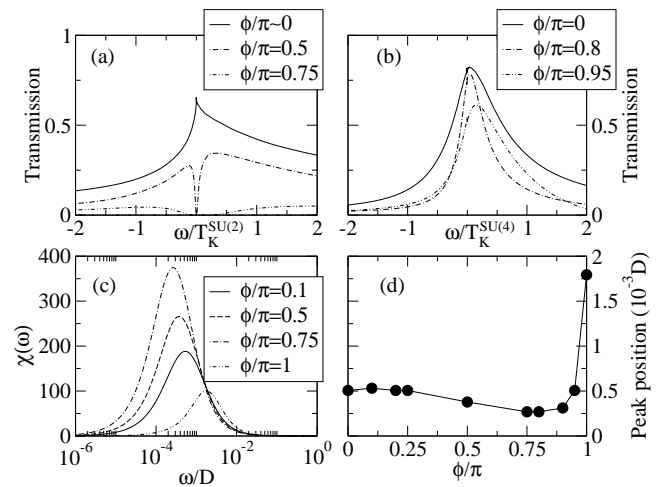}
\caption{\emph{NRG results:} Top panel: Transmission probability versus flux
  for (a) $\veps_{d}=-7\Gamma$, $U_{1}=U_{2}=5D$, $U_{12}=0$,
  and (b) $\veps_{d}=-14\Gamma$, $U_{1}=U_{2}=5D$, and $U_{12}=5D$.
  We set $\Gamma=D/60$.
  (Notice that we do not
  recover the unitary limit of $\mathscr{T}$ for $\phi\approx 0$ (mod $2\pi$) 
  because of the systematic errors introduced in the NRG procedure).
  Bottom panel ($U_{12}=5D$): (c) Spin 
  susceptibility (in an arbitrary unit) in the
  limit of strong interdot interaction. (d) The peak position of the
  susceptibility as a function of the flux $\phi$.}
\label{fig3}
\end{figure}

The value of $\phi_{c}$ is the last ingredient we have to explain.
$\phi_{c}$ marks the crossover between SU(2) Kondo physics to 
the highly symmetric SU(4) Kondo state.
Fortunately, $\phi_{c}$ can be extracted from the
peak position of the spin susceptibility $\chi(\omega)$,
which yields a reasonable estimate of the Kondo temperature.
Figure~\ref{fig3}(c) shows the
evolution of $\chi$ when $\phi$ increases. Remarkably,
when the flux enhances, at some point the position of the peak
moves toward higher frequencies.
The peak position as a function of $\phi$ is plotted in
Fig.~\ref{fig3}(d). We observe that $T_{K}(\phi)$
is almost constant when $\phi$ goes from zero to $\phi_{c}\approx 0.75\pi$.
This fact allows us to establish a criterium for the crossover between
the SU(2) and SU(4) Kondo states in the double QD system.

\section{Conclusion}
\label{conclusions}
We have analyzed the transport properties (in and out of equilibrium) of a
prototypical mesoscopic double-slit interferometer when interactions
play a dominant role. We have shown that crucial differences arise in
the limits of negligible and large the interdot Coulomb interactions.
In the former case, only spin fluctuations matter and each dot develops
a Kondo resonance at the Fermi level independently of the applied
magnetic flux. Due to the interference between these two Kondo 
resonances the linear conductance versus the flux shows 
a series of narrow peaks at $\phi=2\pi$ (mod $2\pi$) 
of unitary height (in units of $2e^2/h$). 
Furthermore, we have found that any deviation from
the Kondo regime (close to the mixed-valence regime) lead to  
dramatic changes in the conductance as a function of
the flux. Interestingly, the nonlinear conductance show the formation
of a dip when $\phi\neq 2\pi$ (mod $2\pi$). A complete suppression of the
electronic transport occurs when the destructive interference condition
takes place $\phi\pi$ (mod $2\pi$).
 
Charge and spin become entangled when the interdot Coulomb interaction
is very large.  Here, the differential conductance
has a zero bias anomaly quenched with increasing flux.
The Kondo state changes its symmetry, from SU(2) to SU(4),
as $\phi$ approaches $\pi$ (mod $2\pi$). 
Since the crossover is not too close to $\phi=\pi$ (mod $2\pi$)
the SU(4) state remains robust to be detected experimentally.
Our geometry requires symmetric couplings to the leads
but not inevitably equal (i.e., we need
$\Gamma_{1L}+\Gamma_{2R}=\Gamma_{1R}+\Gamma_{2L}$).~\cite{note} 
The charging energies $U_{1}$, $U_{2}$, and $U_{12}$ should be of 
the same order (a few meV). Finally, the external flux should correspond 
to a low magnetic field to avoid spin Zeeman splittings in the dot, 
around 10~mT.~\cite{Holl01} All these constraints are experimentally 
accessible with present techniques.~\cite{Weiss,hol03,Holl01}

\section{acknowledgments}
This work was supported by the EU RTN No.
HPRN-CT-2000-00144, the Spanish MECD,
the NSERC, the eSSC at Postech and the SKORE-A. 
K.L.H thanks supports from CIAR, FQRNT, and NSERC.


\begin{thebibliography}{90}
\bibitem{robust}
In particular, 
the SU(4) Kondo state is known to be robust against
small deviations from the perfectly symmetric situation.~\cite{Zar03}
\bibitem{j5}
In a realistic system, the $J_5$ term of Eq.~(\ref{kondo-su4::eq:HKondo:1})
is much smaller than $T_K^{\rm SU(4)}$, preserving the SU(4) state.
\end{thebibliography}
\end{document}